\pdfoutput=1
\documentclass[pdflatex,sn-mathphys-num]{sn-jnl}
\usepackage{graphicx}%
\usepackage{multirow}%
\usepackage{amsmath,amssymb,amsfonts}%
\usepackage{physics}%
\usepackage{mathrsfs}%
\usepackage[title]{appendix}%
\usepackage{xcolor}%
\usepackage{textcomp}%
\usepackage{manyfoot}%
\usepackage{booktabs}%
\usepackage{algorithm}%
\usepackage{algorithmicx}%
\usepackage{algpseudocode}%
\usepackage{listings}%

\usepackage{hyperref}%
\hypersetup{
    colorlinks=true,
    linkcolor=blue,
    citecolor=blue,
    urlcolor=blue,
    breaklinks=true,
    plainpages=false,
    bookmarksopen=true,
    bookmarksnumbered=false,
    bookmarksdepth=5,
    pdfborder={0 0 0}
}

\makeatletter
\def\sectionfont{\subsectionfont}
\def\authemail{}%

\renewcommand\email[1]{%
  \global\advance\emailcnt by 1\relax
  \if@corauemail
    \g@addto@macro\corrauthemail{%
      \setcounter{footnote}{0}%
      \textcolor{blue}{#1};\ %
    }%
  \fi}
\makeatother

\makeatletter
\AtBeginDocument{%
  \def\@pdfborder{0 0 0}%
  \let\@pdfborderstyle\@empty
  \def\@linkbordercolor{1 1 1}%
  \def\@citebordercolor{1 1 1}%
  \def\@urlbordercolor{1 1 1}%
  \def\@filebordercolor{1 1 1}%
  \def\@menubordercolor{1 1 1}%
  \def\@runbordercolor{1 1 1}%
}
\makeatother

\begin{document}

\title[Article Title]{Electrometry of extremely-low frequencies from kHz to sub-Hz with a Rydberg-atom sensor}

\author*[1,2]{\fnm{Aveek} \sur{Chandra}}\email{aveek.chandra@ntu.edu.sg}

\author[3]{\fnm{Narongrit} \sur{Paensin}}\email{iiauthor@gmail.com}

\author[1,2,4]{\fnm{Rainer} \sur{Dumke}}\email{iiiauthor@gmail.com}

\affil[1]{\orgdiv{Centre for Quantum Technologies}, \orgname{National University of Singapore}, \orgaddress{\street{3 Science Drive 2}, \city{Singapore}, \postcode{117543}, \country{Singapore}}}

\affil[2]{\orgdiv{School of Physical and Mathematical Sciences}, \orgname{Nanyang Technological University}, \orgaddress{\street{21 Nanyang Link}, \city{Singapore}, \postcode{637371}, \country{Singapore}}}

\affil[3]{\orgdiv{Department of physics and Materials Science}, \orgname{Chiang Mai University}, \orgaddress{ \city{Chiang Mai}, \postcode{50200}, \country{Thailand}}}

\affil[4]{\orgname{Quantum Sensing Centre}, \orgaddress{\street{2 Science Park Drive}, \city{Singapore}, \postcode{118222}, \country{Singapore}}}







\abstract{Rydberg-atom electric field sensing has shown great potential from near-DC to THz with state-of-the-art measurement metrics realized in sensitivity, phase extraction, multi-band receptivity, etc. While Rydberg-atom sensors have shown exceptional performance in the GHz regime, low-frequency operation has remained challenging because of electric-field-screening in conventional vapor cells, which suppresses externally applied fields. We overcome this limitation by combining auxiliary modulation and lock-in detection with a paraffin-coated vapor cell, and demonstrate an electrode-free, wideband method for sensing frequencies, ranging from 
0.5 Hz to 10 kHz. Our work extends Rydberg-atom sensor's range to VLF, ULF, SLF, ELF and sub-ELF frequency bands. In our method, high state-of-the-art sensitivities have been achieved - 819
$\mu \text{V/cm}/\sqrt{\text{Hz}}$ for 1 Hz, 33 $\mu \text{V/cm}/\sqrt{\text{Hz}}$ for 10 Hz, 10 $\mu \text{V/cm}/\sqrt{\text{Hz}}$ for 100 Hz and 2 $\mu \text{V/cm}/\sqrt{\text{Hz}}$ for 1 kHz.}


\maketitle

\section{Introduction}
\label{sec:intro}
Atomic sensors are central to modern science and technology because atoms of a given species are identical, defect-free, and ultrasensitive to environmental changes. Rydberg atoms, with large DC polarizability and microwave-transition dipole moments, can efficiently measure electric fields over a broad frequency range. However, most demonstrations so far, including those achieving state-of-the-art sensitivities~\cite{Tu2024Approaching, Wu2025EnhancingArrays} and continuous-band detection capabilities~\cite{Liu2022Continuous-FrequencyCell}, have been in the GHz regime, where most conventional wireless communication and sensing occur. Low-frequency electric-field sensing and communication remain valuable for unique applications.

Low-frequency signals are highly penetrating and therefore important for long-range communication, as these wavelengths can propagate over long distances without attenuation, interference, or signal loss. Conventional low-frequency receivers, however, are bulky as resonant antenna size scales inversely with detection frequency, as given by the Chu limit~\cite{Chu1948PhysicalAntennas}. Thus, antennas are typically meter-scale at MHz frequencies and several kilometers at kHz frequencies, making them impractical or unrealizable. Developing low-frequency quantum sensors or receivers that outperform classical receivers is therefore important.

Frequencies in the kHz range are used for underwater-to-air communication because they can penetrate seawater \cite{Latypov2022CompactCommunication}. Low frequencies also enable tracking and localization of buried subsea cables \cite{Szy2019subsea-cables}, and noninvasive testing and characterization of electronic devices and industrial units, as shown for lithium-ion batteries~\cite{Barai2019ACells}. Other applications include studies of electric-fields below 1~Hz, turbulence in atmospheric science \cite{Anisimov2002UniversalAtmosphere}, low-frequency radio astronomy \cite{Bentum2020AAstronomy}, and electromagnetic fields in geophysical \cite{Lseth2006Low-frequencyDiffusion} and biological \cite{Tian2023System-levelReview} systems.

A Rydberg-atom sensor is typically a few centimeters long and can in principle detect electric fields with high accuracy and sensitivity from near-DC to THz frequencies. This wideband capability makes it a practical, field-deployable receiver for RF signals across multiple bands in classical wireless communication and sensing~\cite{Gong2024RydbergSensing}. For example, radar remote sensing of Earth's ecosystem, climate and surface requires detection across multiple bands in different frequency domains. Conventionally, this requires several receivers and electronic systems, each optimized for a narrow frequency window. Instead, a Rydberg-atom sensor can receive multiple bands at different frequencies, shrinking and simplifying cumbersome multi-receiver architectures. Remote sensing of soil moisture with a Rydberg-atom sensor has been shown in Ref. \cite{Arumugam2024remotesens}.

\begin{figure*}[!b] 
	\centering
	\includegraphics[width=0.99\textwidth]
    {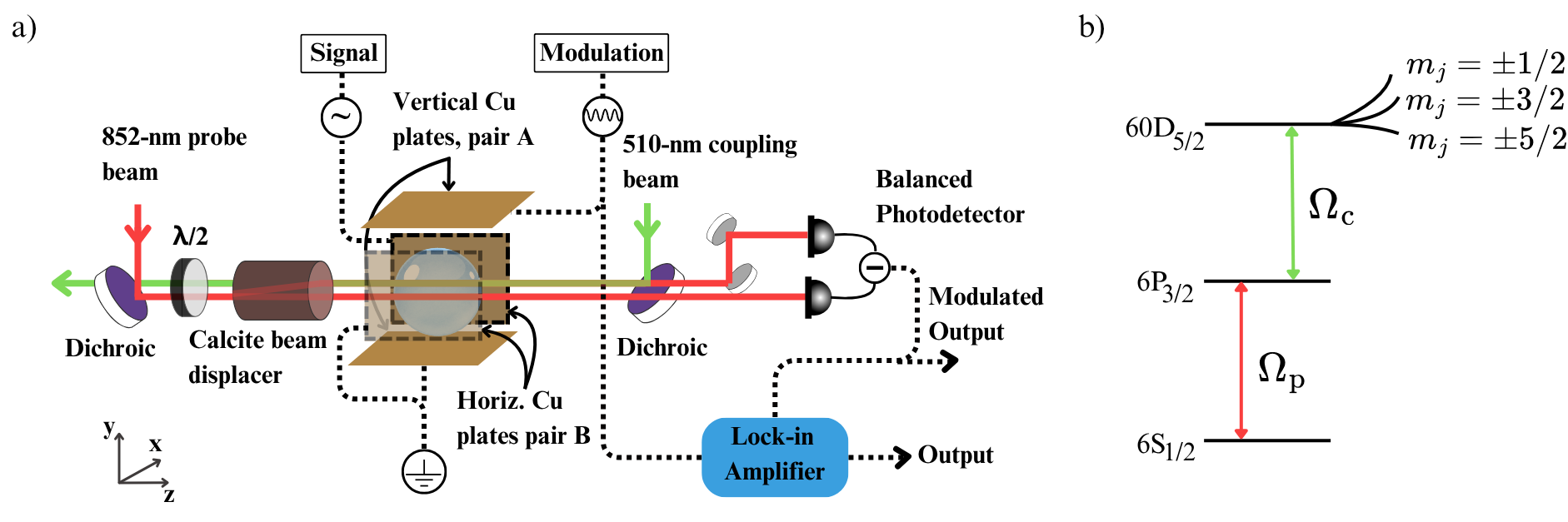} 
	\caption{a) Experimental setup. Probe (852 nm) and coupling (510 nm) beams with effective Rabi frequencies $\Omega_{\text{p}}$ and $\Omega_{\text{c}}$ overlap in a paraffin-coated vapor cell to create two-photon Rydberg excitations to $60D_{5/2}$. A balanced photodetector measures the transmission difference between the probe and an identical reference beam. Auxiliary modulation and signal fields are applied via copper plate pairs, and the lock-in amplifier provides a filtered, demodulated high-SNR output. b) Energy-level diagram for excitation to $60D_{5/2}$. In an electric field, the state splits into $m_j$ sub-levels.}
	\label{fig:schematic} 
\end{figure*}
The challenge in low-frequency Rydberg electrometry is not the intrinsic atomic response but suppression of external electric fields by alkali-induced electric-field screening. This happens as a result of adsorption of a significant fraction of alkali-metal atoms  on the inner SiO$_2$ surface, forming a thin metallic layer that makes the inner cell-wall conductive. When DC or a low-frequency field is applied, free charges in this layer redistribute to generate an opposing field that cancels the applied field. In steady state, the net field inside the cell is therefore zero, so the vapor cell behaves as a `Faraday cage' for DC and low-frequency fields, making their detection impossible. Several studies~\cite{Li2023SuperAtoms, Lei2024HighSensor, Chen25VLF} have shown low-frequency electric-field detection by inserting metal electrodes into the vapor cell, however this complicates the sensor design, reduces sensitivity and overall performance.

One of the first demonstrations of low-frequency electrometry was performed for frequencies in the kHz range with a sapphire vapor cell \cite{Jau2020Vapor-Cell-BasedKHz}. High sensitivity 3.4 $\mu\text{V/cm}/\sqrt{\text{Hz}}$ was achieved for frequencies around $\gtrsim 1$~kHz owing to the passive low-screening operation of sapphire.
However, sapphire vapor cells require specialized fabrication.
In another recent work \cite{Jin2025Heterodyne-lowfreq}, low-frequency sensing was carried out in the kHz range by creating electric-field-induced phase modulation of the probe laser through Rydberg-EIT susceptibility. Subsequently, the signal was extracted via demodulation with optical heterodyne detection. As a conventional vapor cell was used, strong screening factor was reported which limited their measured field-sensitivity to about 1.21 $ \text{mV/cm}/\sqrt{\text{Hz}}$ at 5 kHz. 

In our work, we break the kHz barrier and extend the working point to lower frequencies, down to sub-Hz. We demonstrate electrode-free sensing of extremely-low frequency electric fields ranging from 0.5 Hz to 10 kHz, covering sub-ELF, ELF, SLF, ULF and VLF frequency bands.  By using a paraffin-coated cell, an auxiliary modulation field and lock-in detection technique, high sensitivities have been achieved and the values reported in Fig. \ref{fig:SNR_vs_freq}(a).

The paraffin-coating significantly slows the electric-field-screening dynamics by delaying the buildup of the screening charge. The auxiliary modulation field periodically reverses the internal field before the screening charge distribution has reached equilibrium. This dynamic field-refreshing process, together with lock-in detection, transfers the low-frequency signal to a narrow detection band, allowing weak electric fields at very low frequencies to be measured with a high signal-to-noise ratio.

\section{Methods and Results}\label{sec:result}
The experimental schematic and the relevant energy levels of a Cs atom are presented in Fig.~\ref{fig:schematic}(a) and (b) respectively.
We have used a spherical paraffin-coated vapor cell (pyrex) at room temperature in the experiment, the relevant details of the cell can be found in Appendix \ref{sec:vp_cell}.
A probe laser at 852 nm, out-coupled from a polarization-maintaining fiber, is collimated and split into two beams - a probe beam with waist 0.3 mm and an identical reference beam. They propagate in parallel through the cell and finally collected on a balanced photodetector (PD) where the probe signal is subtracted from that of reference to obtain the transmission signal.
A coupling laser at 510 nm, out-coupled from a polarization-maintaining fiber, is collimated to a beam waist of 0.45 mm. It counter-propagates and overlaps with the probe beam inside the vapor cell. Both probe and coupling beams have collinear polarization along the y-axis. The probe laser drives the D2 transition of Cs represented as $6\ket{S_{1/2}}\rightarrow 6\ket{P_{3/2}}$ with effective Rabi frequency $\Omega_{\text{p}}=2\pi\cross 25.6$ MHz (optical power $\sim$ 200$\mu$W). The coupling laser drives the Rydberg excitation to $n=60$ represented as $6\ket{P_{3/2}}\rightarrow 60\ket{D_{5/2}}$ with effective Rabi frequency $\Omega_{\text{c}}=2\pi\cross 1.7$ MHz (optical power $\sim$ 42 mW).

Under two-photon resonance, Rydberg-EIT transparency peak appears in the transmission signal when the probe frequency is scanned over resonance. A balanced detection mode is used to remove the Doppler-broadened absorption background, and also to cancel out any intensity noise of the probe laser.
In the experiment, both probe and coupling lasers are frequency-locked to a reference cavity - a dual-axis, cubic, ultra-low expansion (ULE) cavity, with very high-finesse. The details on the setup and laser-locking methods can be found in Appendix \ref{sec:laser_stabilize}.
\subsection{Application of DC E-field}\label{sec:appl_Efield}
\begin{figure}[!t] 
	\centering
	\includegraphics[width=0.91\textwidth]{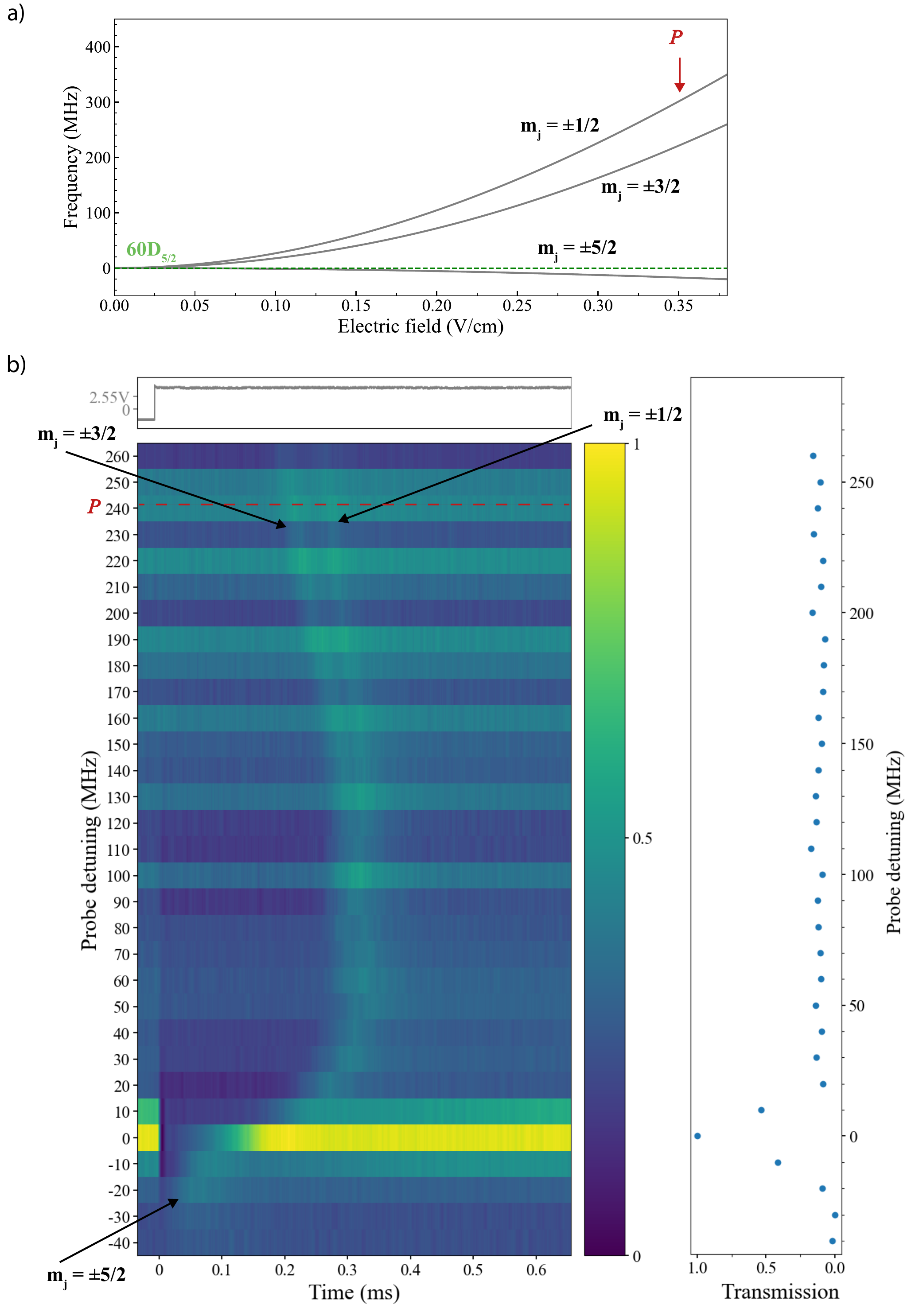} %
    \caption{(a) Calculated Stark spectrum of the 60D$_{5/2}$ state for fields from 0 to 380 mV/cm, showing the $m_j$-dependent Stark shift. (b) At $t=0$, a DC bias field of amplitude 354 mV/cm ($\text{V}_p=2.55$ V) is switched on and the transient response is recorded at different probe detunings in a paraffin-coated cell. The resulting 2D color map shows the three split branches corresponding to $m_j=\pm 5/2,\pm 3/2,\pm 1/2$, with most atoms in $\pm 3/2,\pm 1/2$. The red dashed line `$P$' marks $\delta_p=243$ MHz, used for sensor operation. Right: Rydberg-EIT spectrum in the absence of electric field.}
    \vspace{-10pt}
	\label{fig:Stark_3Dplot}
\end{figure}

Rydberg states experience quadratic energy (Stark) shift ($\Delta f$), in the presence of electric field ($E$). This is given by:
\begin{equation}
\Delta f=-\frac{\alpha}{2} E^2
\label{eqn:Stark_shift}
\end{equation}
where $\alpha$ is the DC polarizability of the Rydberg state.
Fig.~\ref{fig:Stark_3Dplot}(a) shows the energy shift experienced by Rydberg state $60D_{5/2}$ when DC electric field is applied. With increase in electric field strength, the degeneracy in $m_j$ sublevels is lifted that results in splitting and $m_j$ dependent Stark shifts. The calculated polarizabilities (using ARC Rydberg calculator~\cite{SibalicARC2017}) of the sublevels $m_j=\pm1/2, \pm3/2, \pm 5/2$ are -4985 MHz cm$^2/$V$^{2}$, -3624 MHz cm$^2/$V$^{2}$, 281 MHz cm$^2/$V$^{2}$ respectively.

In the experiment, the electric fields are applied to the paraffin-coated cell, via pairs of copper plates, A in vertical and B in horizontal direction as shown in Fig.~\ref{fig:schematic}(a). The plates have dimensions of a square of length 70 mm and thickness of 2 mm. The vertical plates (pair A) are separated by 72 mm while horizontal plates (pair B) are apart by 76 mm. 
The paraffin-coated, spherical vapor cell with a radius of 15 mm, is placed at the center in between the pairs of plates such that the electric field is uniform across the dimension of the cell.
\begin{figure}[!b] 
	\centering
	\includegraphics[width=0.9\textwidth]{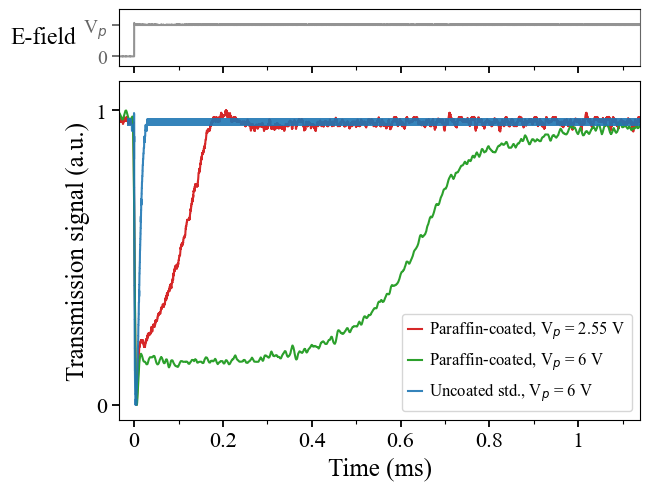} %
	\caption{Transient diagnostic of electric-field screening in vapor cells. The balanced photodetector output is recorded directly after a DC electric-field step is applied at (t=0). The initial drop occurs because the Stark shift moves the Rydberg-EIT resonance away from the fixed laser frequency. The subsequent recovery occurs as wall charges screen the internal electric field and return the EIT resonance toward its original position. The recovery time defines the screening time constant $\tau$. For standard (uncoated) cell, $\tau=10\text{ }\mu$s is independent of the magnitude of $\text{V}_p$. For paraffin-coated cell, $\tau =$ 0.1 (0.6) ms for $\text{V}_p=$ 2.55 (6) V, which in terms of field amplitude = 354 (833) mV/cm.}
	\label{fig:signal_decay} 
\end{figure}

At first, we study the effect of electric-field-screening in paraffin-coated cell relative to an uncoated (standard) cell. With both probe and coupling lasers frequency-locked to the reference cavity, we record the transmission signal at two-photon resonance. The transient responses for the standard (uncoated) cell and the paraffin-coated cell under the application of E-field with amplitude of 833 mV/cm is shown in Fig.~\ref{fig:signal_decay}. When the E-field is turned on at time $t=0$, transmission signal rapidly drops to zero. This is because the two-photon resonance condition changes causing the Rydberg-EIT transmission peak to be shifted in frequency. In fact, the rate of decay of transmission signal is given by the bandwidth of the balanced photodetector.
Now, because of the screening-effect from the free-surface charges, the E-field inside the vapor cell is slowly reduced to zero over some characteristic time, shifting the Rydberg-EIT peak back to its original resonant frequency. We exploit the slow electric-field-screening dynamics of the paraffin-coated cell to enable low-frequency electric-field sensing.

The characteristic time over which the signal recovers to 0.5 of its initial value, is defined here as the screening-time constant $\tau$.
For standard (uncoated) cell, the transient signal (in blue) rapidly recovers, following an exponential profile, to its original value, $\tau=10 \text{ }\mu$s. However, for paraffin-coated cell, the transient signal recovers at a slower rate, and the rate even depends on the applied field amplitude. $\tau =$ 0.1 (0.6) ms for $\text{V}_p=$ 2.55 (6) V, which in terms of field amplitude is 354 (833) mV/cm.
The different electric-field screening times for uncoated and paraffin-coated cells can be qualitatively explained as follows.

For standard (uncoated) galss cell, the adsorption of cesium atoms on the inner glass surface creates an effective thin metallic film with conductivity that is orders of magnitude higher than the conductivity of glass. This allows rapid redistribution of surface-free charges similar to that of a metallic conductor giving an electric-field screening time $\tau = 10 \text{ }\mu s$. Assuming a simple RC model for dipolar surface-charge distribution formed by the alkali-induced conducting layer on the inner surface of the spherical-uncoated cell, one can estimate (order-of-magnitude) the sheet resistance, $R_s \sim 30$ M$\Omega$ (details in Appendix \ref{sec:vp_cell}).

Compared with bare glass, a paraffin-coated wall provides an electrically insulating and chemically different surface on which alkali adsorption and surface-charge transport are expected to be strongly reduced. Alkali diffusion into paraffin coatings and the formation of alkali aggregates or whiskers under some conditions have been reported previously~\cite{Balabas2007, Atutov2017}, and light-induced changes in alkali-vapor-cell coatings have also been observed~\cite{Hibberd2013}. These results indicate that a paraffin-coated wall should not be treated as an ideal, charge-free insulator. Nevertheless, the density and mobility of charges available for screening are expected to be much lower than on an alkali-covered bare-glass wall. This interpretation is consistent with our measured screening transients: the uncoated cell screens within approximately 10 $\mu$s, whereas the paraffin-coated cell screens on 0.1-0.6 ms timescales. The increase of the apparent screening time with field amplitude suggests that the rate at which screening charge can be redistributed on or within the coated surface is limited. A simple charge-supply picture would then give a longer screening time for larger fields, since the charge density required to screen the applied field $E$ scales approximately as $\sigma_{\mathrm{screen}} \sim \epsilon E$, where $\epsilon$ is the effective
electric permittivity of the space. We emphasize that this is a qualitative model; the microscopic charge-transport pathway in the coated cell is not directly measured here.

To move beyond acquisition on two-photon resonance and identify an optimal operating point $P$, we study the transmission signal at different probe detunings under an applied E-field. In the experiment, the probe detuning is varied with a fibered acousto-optic frequency shifter. The transient recorded at each detuning is compiled into the 2D color map shown in Fig.~\ref{fig:Stark_3Dplot}(b). Figure~\ref{fig:Stark_3Dplot}(a) shows the calculated Stark spectrum of the 60D$_{5/2}$ state, which splits into its $m_j$ branches as the field amplitude increases. Since each branch has a different DC polarizability, the operating point matters. The point `$P$' corresponds to an applied field of 354 mV/cm. In the experiment, both the laser polarization and the applied field are along the vertical $y$ direction, so most atoms populate the $m_j=\pm 1/2,\pm 3/2$ branches, as seen in Fig.~\ref{fig:Stark_3Dplot}(b). The red dashed line marked `$P$' corresponds to a probe detuning of $\delta_p=243$ MHz, used later for sensor operation.

First, direct sensing of low-frequency electric fields, without an auxiliary field, was performed. The field was applied to the atoms via a pair of copper plates, and sensing was demonstrated only for frequencies $\geq 500$ Hz and amplitudes $\geq 100$ mV/cm as shown in Appendix \ref{sec:direct_lowfrsens}.

\subsection{Sensor operation with auxiliary modulation}
A sensor's sensitivity is proportional to slope of its response curve at the operating point $P$. Here, the response is induced by Stark shift (see Eq.~\ref{eqn:Stark_shift}) of the state (or branch) $m_j$, with polarizability $\alpha_j$, under application of electric field amplitude $E$:
\begin{equation}
\text{Sensitivity} \propto |\dv*{\Delta f}{E}|_P = \alpha_j E
\label{eq:E_sense}
\end{equation}

Sensitivity can be enhanced if operation point $P$ has high slope, which requires both a highly polarizable Stark branch and a larger field amplitude, provided the EIT line is not overly broadened or saturated. This is achieved in our method by using an auxiliary modulation field $E_{\text{aux}}$ and choosing a probe detuning $\delta_p$ so that operation region lies on more polarizable branches $m_j=\pm1/2$ and $m_j\pm3/2$ over the weakly polarizable $m_j=\pm5/2$ branch (see calculated polarizabilities in Sec\ref{sec:appl_Efield}). From Stark map in Fig.~\ref{fig:Stark_3Dplot}(a), we see that a larger amplitude $E_{\text{aux}}$ requires a larger positive detuning $\delta_p$ to sense weak fields (high-sensitivity domain), whereas a smaller $E_{\text{aux}}$ requires a smaller detuning, close to two-photon resonance to sense relatively strong fields (low-sensitivity domain). Therefore, operation in both domains can ensure a large linear dynamic range of the sensor.

In this work, we focus on the former, high-sensitivity domain and use $E_{\text{aux}}=354$ mV/cm and $\delta_p=243$ MHz for sensing. The optimized point $P$ was obtained by scanning over a range of $E_{\text{aux}}$ and $\delta_p$ and then selecting maximum fitted lock-in response (details in Appendix \ref{sec:2D_grid}). Although, the operating point $P$ (marked in Fig.~\ref{fig:Stark_3Dplot}(b)) is determined primarily by the applied auxiliary modulation field, small residual DC electric fields may arise from coupling-light-induced surface charges on the vapor-cell wall.

Residual DC fields generated by coupling-light-induced surface charge are an important concern in Rydberg-EIT electrometry. Such fields have been observed in glass vapor cells under visible coupling-beam illumination~\cite{Patrick2025}, and laser-induced DC fields have also been used deliberately in sapphire cells for Rydberg sensing~\cite{Zhang2026}. In the present paraffin-coated cell, We do not observe any resolvable Stark splitting of the Rydberg-EIT resonance at zero applied electric field [Fig. \ref{fig:Stark_3Dplot}(b), right panel]. This indicates that any static residual field in the probed region is below our spectral resolution and is small compared with the deliberately applied auxiliary field used at the operating point $P$. We determine $P$ empirically by scanning $E_{\mathrm{aux}}$ and $\delta_p$ under the same optical conditions used for sensing; therefore, any unresolved static offset is automatically included in the calibration.

\begin{figure*}[!ht] 
	\centering
	\includegraphics[width=0.98\textwidth]{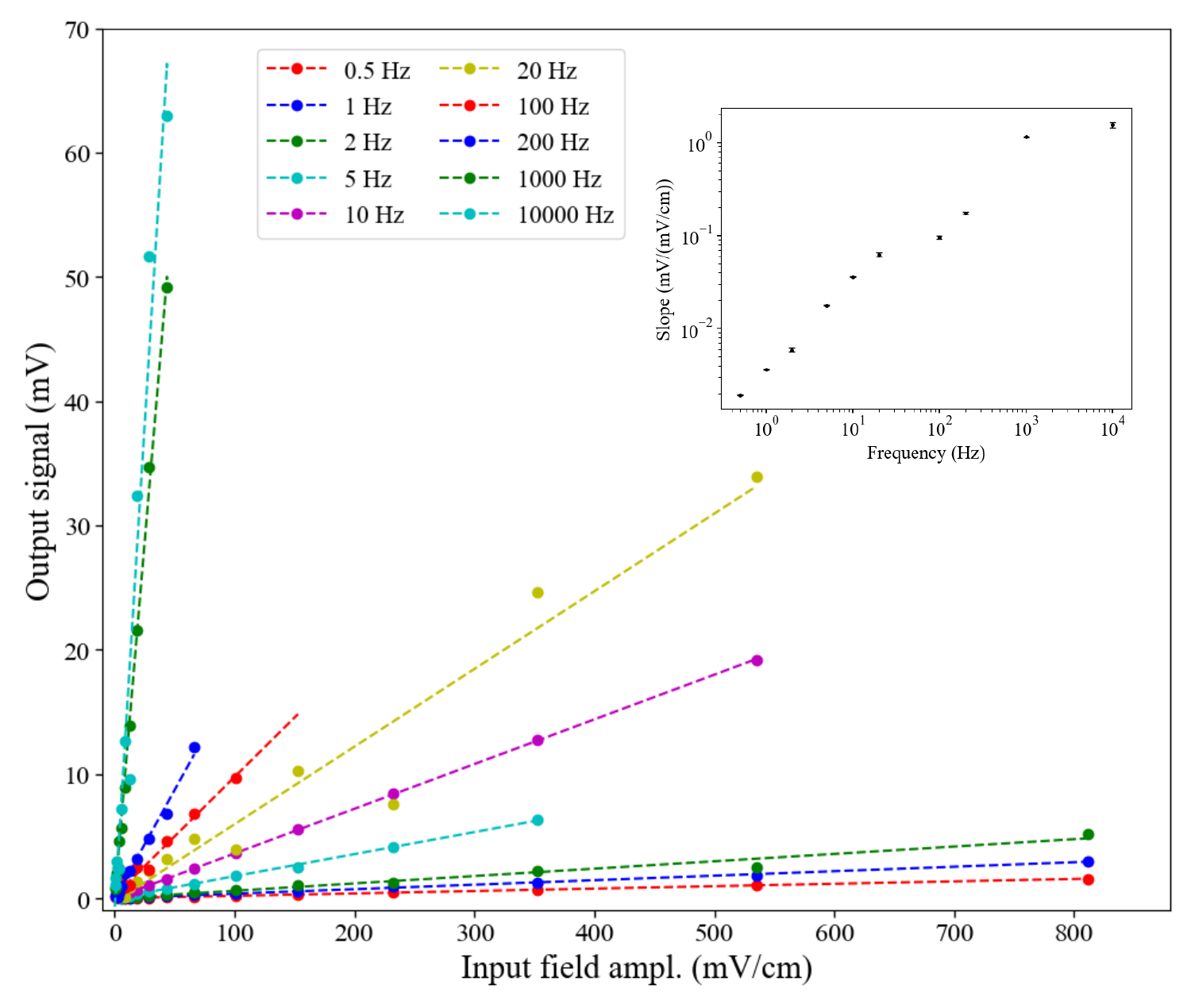}
	\caption{Output signal versus input field amplitude in the linear regime for different low frequencies. Each dashed line is a linear fit to the data for the sensed frequency. Inset: slope $m(f)$ from the linear fit, plotted versus signal frequency, giving the sensor responsivity.}
	\label{fig:sensing_outvsinp}
\end{figure*}

The modulation frequency $f_{mod}$ is chosen so that the polarity of the auxiliary field flips when electric-field-screening begins to affect the atoms inside the cell. From Fig.~\ref{fig:Stark_3Dplot}(b), the $m_j=\pm 3/2,\pm 1/2$ branches cross the dashed line $P$ at about 0.25 ms, corresponding to 4 kHz, which sets the lower limit on the modulation frequency. Its value is set empirically by maximizing the signal-to-noise ratio. For sensing frequencies from 0.5 Hz to 200 Hz, we use $f_{mod}=7.9$ kHz. For 500 Hz and 1 kHz, $f_{mod}=27.9$ kHz is used, while for 10 kHz, $f_{mod}=87.9$ kHz is used.

In the experiment, the auxiliary modulation field $E_{\text{aux}}$ and the signal field $E_{\text{sig}}$ are applied to copper plate pairs A (vertical) and B (horizontal) respectively, as shown in Fig.~\ref{fig:schematic}(a). The modulated output of the balanced photodetector (PD) is sent to one input of the lock-in amplifier (LIA), while the modulation field $E_{\text{aux}}$ serves as the reference at the other input, as shown in Fig.~\ref{fig:schematic}. This enables synchronous demodulation at $f_{mod}$; after appropriate filtering, the output provides the sensor signal. In the experiment, the low-pass filter cutoff is chosen as 1.265$f$, where $f$ is the frequency being sensed.

With the sensor operated at the point $P$, electric fields from 0.5 Hz to 10 kHz were detected, with both frequency and amplitude measured for an input field of 43.7 mV/cm, as shown in Fig. \ref{fig:lowfreq_sine} in Appendix \ref{sec:demod_out}. The measurements were repeated over input amplitudes from 0.5 $\mu$V/cm to 1 V/cm. The output begins to saturate as the input amplitude increases, and the saturation point depends on frequency. Figure~\ref{fig:sensing_outvsinp} shows the measured output signal in the linear regime as a function of input amplitude for all frequencies. The slope $m(f)$ from the linear fit for each frequency, i.e. the sensor responsivity, is plotted versus signal frequency in the inset.

The variation of responsivity with signal frequency can be attributed to residual electric-field-screening. The auxiliary field at frequency $f_{mod}$ does not eliminate screening completely, but mitigates it. As a result, the field experienced by the atoms before the auxiliary field switches is smaller than the applied, unscreened field. This residual screening therefore affects lower-frequency, longer-duration signals more strongly than higher-frequency, shorter-duration signals, accounting for the increase in responsivity $m(f)$ as the signal frequency rises from 0.5 Hz to 10 kHz.

\subsection{Sensitivity spectrum}
\label{sec:sensitivity_spec}
The electric-field sensitivity at a given signal frequency can be estimated by identifying and characterizing the relevant noise sources in the experiment. The main contributors are laser intensity fluctuations and power drift, laser polarization drift, and noise from the balanced photodetector (PD) and lock-in amplifier (LIA). From Fig.~\ref{fig:schematic}(a), any polarization noise or drift is converted into intensity noise or drift.

Noise from the PD and laser intensity is measured together by recording the output of the balanced-PD with a spectrum analyzer (Fig. \ref{fig:PD_noise_meas}(a) in Appendix \ref{sec:noise-analysis}). From this measurement, the relevant noise spectral density $\text{NSD}_{\text{PD}}$ at the modulation frequency $f_{mod}$ is obtained. The other contribution is the intrinsic noise of the LIA, which is inevitably added to the demodulated output. This flicker noise is intrinsic to the device and is specified by the LIA input voltage noise density $\text{NSD}_{\text{LIA}}$ in the device manual. The two separate noise contributions originating from PD and LIA are illustrated in Fig. \ref{fig:PD_noise_meas}(b) in Appendix \ref{sec:noise-analysis}). For uncorrelated noise sources, the total noise spectral density is
$\text{NSD}_{\text{total}}=\sqrt{\left(\text{NSD}_{\text{PD}}\right)^2 + \left(\text{NSD}_{\text{LIA}}\right)^2}$.\\
Using the responsivity $m(f)$ discussed above, the field sensitivity at frequency $f$ is estimated as $\text{S}_E = \text{NSD}_{\text{total}}/m(f)$.
\begin{figure}[!b] 
	\centering
	\includegraphics[width=0.95\textwidth]{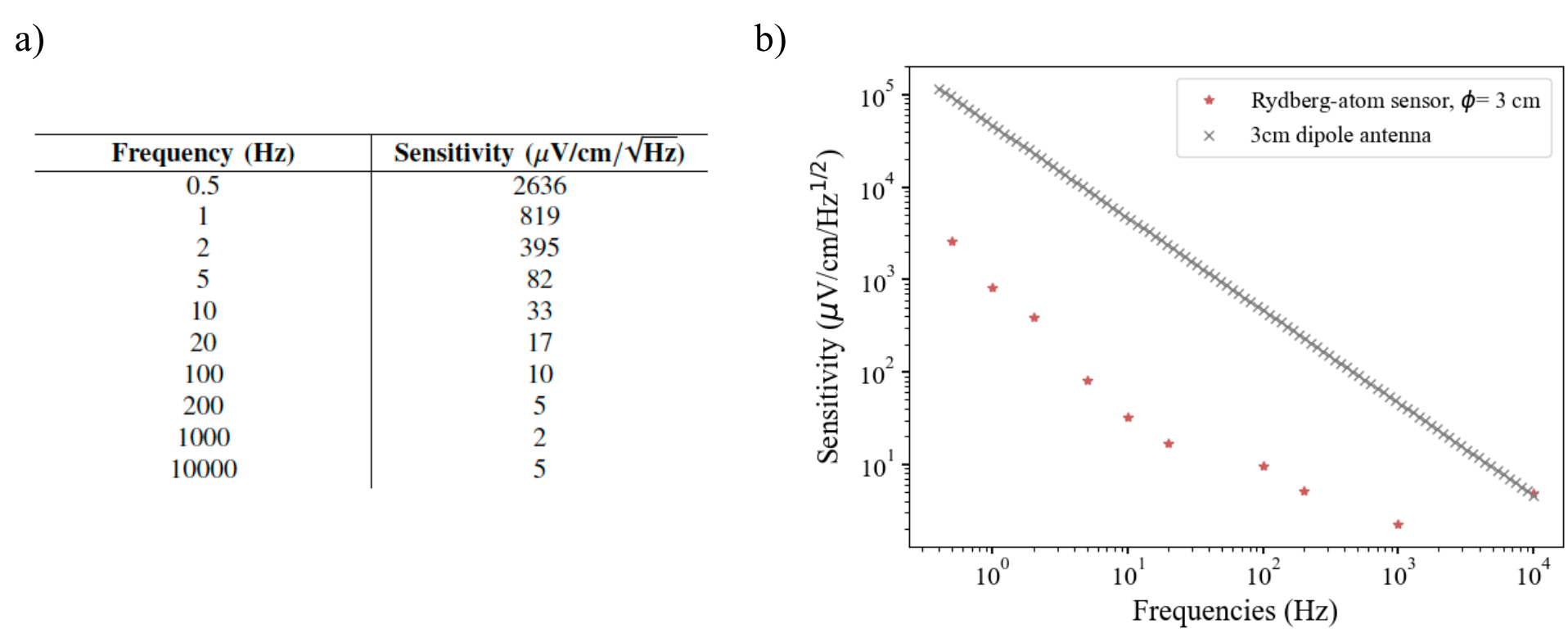}
	\caption{(a) Measured sensitivities of our Rydberg-atom sensor at various frequencies. (b) These sensitivities for vapor-cell with diameter $\phi=3$ cm (red), compared with the theoretically estimated field sensitivity of a classical receiver, a 3 cm dipole antenna (gray), at different frequencies. The Rydberg sensor shows about one to two orders of magnitude better sensitivity at extremely low frequencies. For frequencies $\gtrsim 1$ kHz, the two sensitivities begin to approach each other.}
	\label{fig:SNR_vs_freq}
\end{figure}

The sensitivities achieved with the Rydberg-atom sensor using a vapor cell of diameter $\phi=3$ cm are plotted in red in Fig.~\ref{fig:SNR_vs_freq} as a function of signal frequency. For comparison, the theoretically estimated field sensitivity of a classical receiver \cite{Meyer2021assessment} of the same size - 3 cm dipole antenna, is shown in gray. The Rydberg-atom sensor is about one to two orders of magnitude more sensitive than the classical receiver at extremely low frequencies. For frequencies $\gtrsim 1$ kHz, the sensitivities of the classical and quantum sensors begin to approach each other.

Several techniques already implemented in the experiment lower the spectral noise floor. Frequency noise of the probe and coupling lasers, which is converted into intensity noise by the atomic spectroscopy, is mitigated by locking the laser frequencies to a high-finesse ULE cavity. Using a large probe detuning, $\delta_p=243$ MHz, far from resonance, further reduces frequency-induced intensity noise. Finally, lock-in detection restricts the measurement to a narrow bandwidth and filters out broadband spectral noise. At present, probe-laser intensity fluctuations and power drifts primarily limit the sensor sensitivity. Ongoing noise-mitigation work is expected to improve it further.

\section*{Outlook}
We report here Rydberg-atom based wideband sensing of low frequencies ranging from sub-Hz to 10 kHz with a paraffin-coated cell. The delayed electric-field screening in paraffin-coated cell, combined with dynamic-field reversing provided by the auxiliary modulation field, creates a temporal window during which the atoms, before being shielded from the external electric field, can probe and sense the applied electric field. Our approach is readily extendable to higher-frequency electric fields, covering the range from a hundred kilohertz up to about 10 MHz, provided a high-bandwidth balanced photodetector and a lock-in amplifier suitable for high-frequency demodulation are used. A suitable operating point, $P$, must also be selected.

Ongoing efforts are focused on developing a compact, integrated Rydberg-atom sensor suitable for future field deployment. Such a system will require the paraffin-coated cell, compact collimators, balanced photodetector, and field plates to be packaged into a fiber-coupled sensor head. The present ULE-cavity laser-locking system can be miniaturized using a compact cavity or transfer-lock architecture.  Portable Rydberg atomic antennas have shown compact operation with significantly reduced system size, weight, and complexity. For example, Ref.~\cite{Wu2026} implements integrated laser-frequency-stabilization scheme together with a waveguide-integrated resonator that enhances the RF field used for atomic heterodyne detection. For outdoor low-frequency operation, however, additional measures are necessary to suppress environmental electric-field pickup, ground loops, and slow drifts arising from thermal and polarization instabilities.

Paraffin-coated cells, widely used for all-optical magnetic-field sensing, are shown here to be excellent candidates for electric-field sensing, particularly in the low-frequency domain. This opens the possibility of simultaneous, correlated sensing of electric and magnetic fields in biological specimens, novel materials, and geological studies. It may also open new frontiers in low-frequency radio astronomy and aid the search for exotic particles and fields via global sensor networks such as GNOME~\cite{Afach2018CharacterizationGNOME}.
\newcounter{mainfig}
\setcounter{mainfig}{\value{figure}}
\begin{appendices}
\setcounter{figure}{\value{mainfig}}
\renewcommand{\thefigure}{\arabic{figure}}
\section{Laser-frequency stabilization and device photos}
\label{sec:laser_stabilize}
In atomic spectroscopy experiments, laser frequencies typically need to be locked to a frequency reference for achieving high-degree of frequency stability. In our case, this is provided by a cubic, dual-axis, ultra-low expansion (ULE) cavity placed in high vacuum environment. The ULE cavity (from Stable Laser Systems) has two axes, each with mirrors customized with respective high-reflectivity coatings at 510 nm and 852 nm. The specifications are: FSR $=3$ GHz, finesse $>250000$ for 852 nm axis and finesse $=50000-150000$ for 510 nm axis.
\begin{figure*}[!t] 
    \vspace{-6pt}
	\centering
	\includegraphics[width=0.92\textwidth]{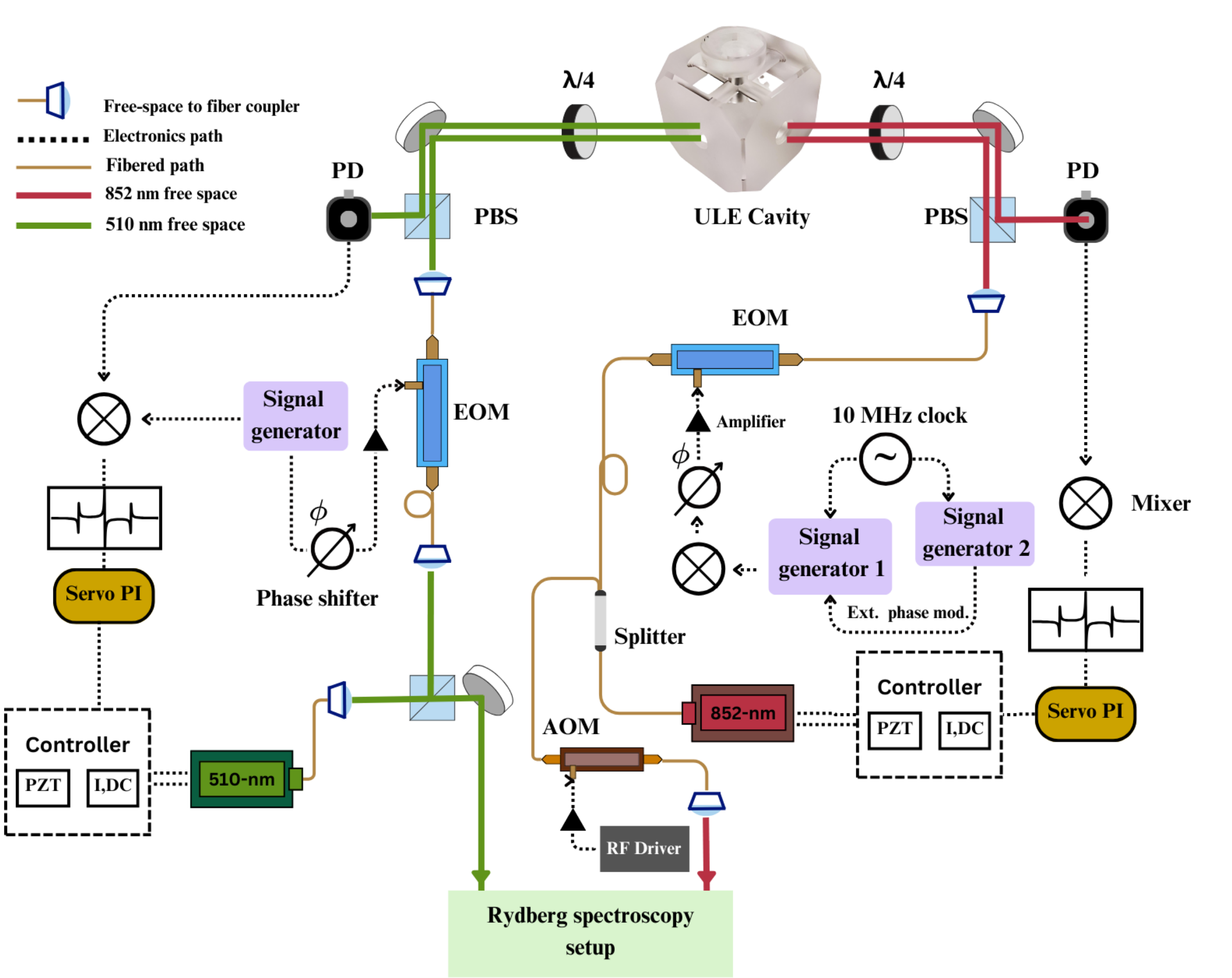} 
	\caption{Setup for frequency-stabilization and beam preparation. A fraction of probe and coupling lasers are sent to ULE cavity for frequency locking via PDH scheme. The rest of their power, after preparation, are sent to Rydberg spectroscopy setup for sensing.}
    \vspace{-2pt}
	\label{fig:setup} 
\end{figure*}

The full experimental schematic can be understood by combining Fig~\ref{fig:setup} and Fig~\ref{fig:schematic} from the main text.
A fibered-laser at 852 nm (probe) is split into two outputs by a fibered splitter (coupler). One output is sent to a fibered acousto-optic frequency shifter (AOM) to set any desired probe detuning before being out-coupled from fiber into free-space for Rydberg-atom sensing. The other output is again divided between saturated absorption spectroscopy setup (for atomic-referencing) and laser-frequency stabilization setup. Similarly, a fiber-coupled laser at 510 nm (coupling) is divided into two parts - one for Rydberg-atom sensing and the other for laser frequency stabilization.

Pound-Drever-Hall (PDH) scheme is implemented to lock the laser frequencies to the reference ULE cavity. The coupling beam is phase-modulated by a fibered-EOM (from Jenoptik) that creates frequency sidebands about a fixed frequency (typically a few MHz). The modulated beam with polarization appropriately aligned, is coupled into the ULE cavity. The beam coupled out of the cavity is detected by low-noise photodiode (PD) whose output signal is then demodulated to produce the error signal that is fed back, via servo-PID controller, to the laser. The fast feedback is sent to laser current while the slow feedback goes to laser piezo (PZT) system.

The 510-nm coupling laser is PDH-locked to a cavity mode, which has a frequency offset of approximately 40 MHz from the Rydberg transition, $6P_{3/2}\rightarrow60D_{5/2}$.
For the 852-nm probe laser, we found the cavity-transmission peak is far off in frequency, approximately 1 GHz away, from the two-photon resonance. Offset-sideband locking is employed to lock one modulation sideband to a cavity mode while maintaining the carrier on the Cs D2 transition. The carrier is subsequently frequency shifted by an acousto-optic modulator, and the resulting probe detuning, together with coupling-frequency offset (40 MHz), sets the desired two-photon detuning.
As shown in Fig.~\ref{fig:setup}, a fiber-coupled EOM generates the required offset sideband using the RF signal from signal generator 1. Additional PDH phase-modulation sidebands are then imposed around this offset sideband using an external phase-modulation signal from signal generator 2. The remainder of the frequency-locking architecture is identical to that used for the coupling laser. The probe detuning used for the operating point $P$, $\delta_p=243$ MHz is therefore defined relative to the measured Rydberg-EIT resonance.
\begin{figure}[!h] 
    \vspace{-4pt}
	\centering
    \makebox[\textwidth][c]{%
	\includegraphics[width=0.93\textwidth]{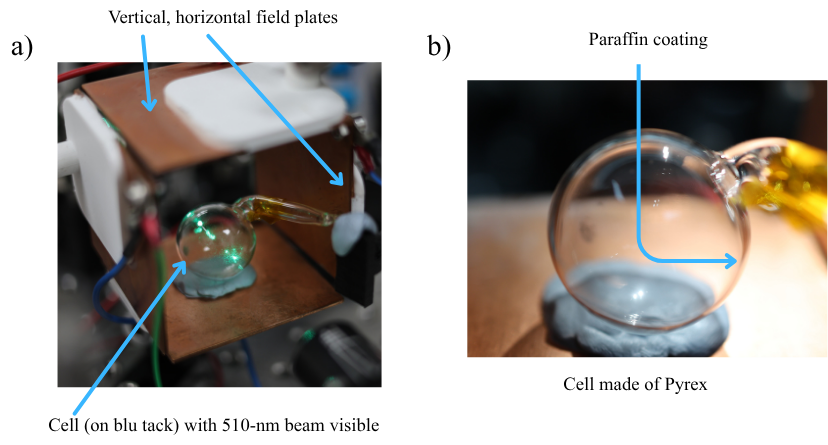}
    }
	\caption{Photographs of paraffin-coted cell (a) on blu tack with 510-nm beam shining through it; mounted between copper plates. Plates have side length 70 mm and thickness 2 mm. The vertical and horizontal plate separations are 72 mm and 76 mm, respectively. (b) in close-up, cell radius=15 mm; the paraffin coating is transparent (not visible) and located on the inner wall of the sealed cell.}
	\label{fig:photos_device}
\end{figure}

In Fig. \ref{fig:photos_device}, we provide photographs showing our paraffin-coated Cs Pyrex vapor cell
mounted between copper field plates, together with a close-up view of the spherical cell.

\section{Vapor cell characteristics}
\label{sec:vp_cell}
The vapor cell used in this work is a custom spherical paraffin-coated Pyrex Cs cell fabricated in the group of  Prof. A. Weis, Université de Fribourg. The cell radius is 15 mm and the cell is operated at room temperature. The same type of paraffin-coated spherical Cs cell has previously been used in cavity-enhanced atomic magnetometry \cite{Crepaz2015}, where the measured ground-state relaxation rate $\Gamma_{\text{relax}} 
=2\pi\times 4.4$ Hz at room temperature, demonstrated the high anti-relaxation quality of the coating. Figure \ref{fig:photos_device}(a) show a photograph of the actual cell (on blu tack) mounted in between electric-field-plates.  
Figure \ref{fig:photos_device}(b) shows close-up of our cell indicating Pyrex wall and paraffin coating. The paraffin layer is transparent and located on the inner surface of the sealed cell.

\textbf{Sheet resistance estimation:}
The estimate based on a surface-RC model of the conducting alkali layer formed on the inner wall of the uncoated cell. The screening of a uniform external field corresponds to the dipolar, ($l$=1), charge redistribution mode on the spherical inner surface. 

The experimentally measured screening time is
$\tau \approx 10~\mu\mathrm{s}$.
For a spherical cell of radius $a=15$ mm, the effective capacitance of the dipolar surface mode is approximately
$C_{\mathrm{eff}} \approx \frac{2\pi \epsilon_0 a}{3}.$
Using a = $1.5\times 10^{-2}$ m and $\epsilon_0 = 8.85\times 10^{-12}$ F/m gives
$C_{\mathrm{eff}} \approx \frac{2\pi(8.85\times10^{-12})(1.5\times10^{-2})}{3} \approx 3\times10^{-13}~\mathrm{F}.$
The sheet resistance is then estimated from the time constant
$\tau \approx R_{s} C_{\mathrm{eff}},$
so that
$R_{s} \approx \frac{10\times10^{-6}}{2.8\times10^{-13}} \approx 3\times10^{7}~\Omega.$
\section{Finding optimal operating points}
\label{sec:2D_grid}
\begin{figure}[!b] 
\vspace{-6pt}
	\centering
	\includegraphics[width=0.63\textwidth]{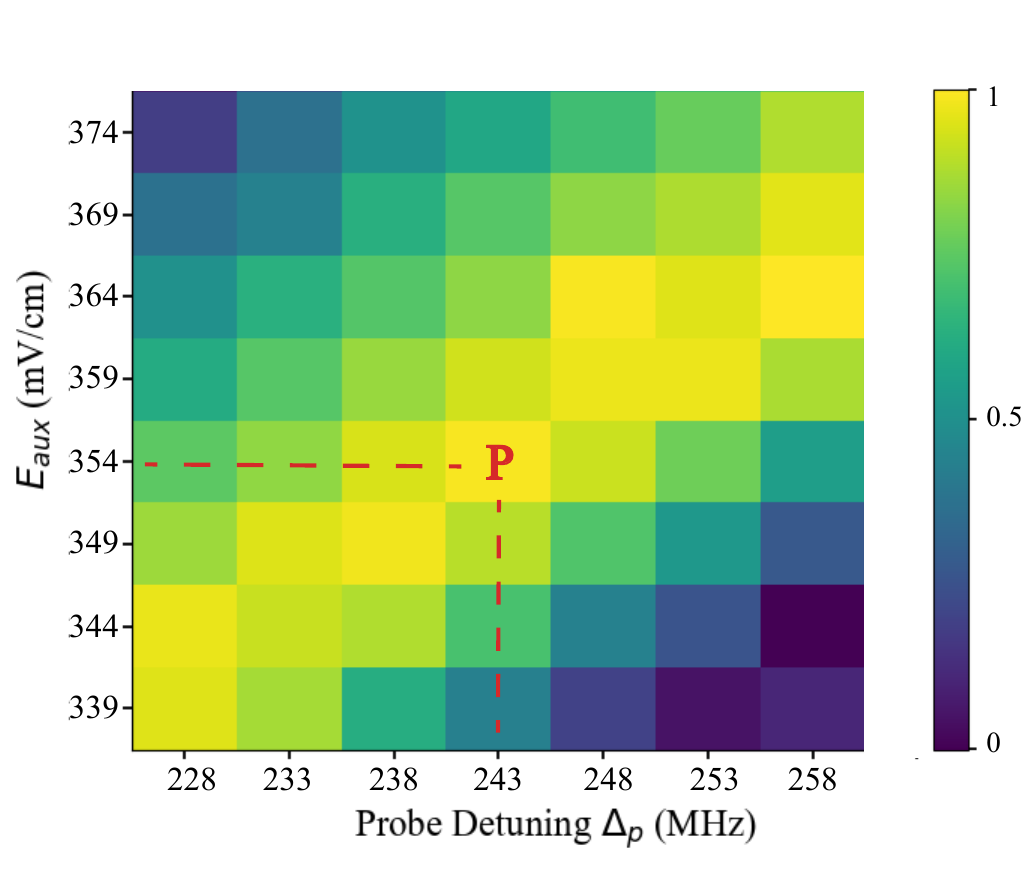} 
	\caption{ 2D grid for a range of auxiliary field amplitudes $E_{\text{aux}}$ and probe detunings $\delta_{p}$. The color scale represents fitted sine amplitudes at the output for an input field of frequency 10 Hz and amplitude 100 mV/cm. The chosen operating point (square marked with `P') is: $E_{\text{aux}}=354$ mV/cm, $\delta_{p}=243$ MHz.}
    \vspace{-8pt}
	\label{fig:scanned_grid} 
\end{figure}
The sensor operation depends largely on amplitude and frequency of auxiliary modulation field$E_{\text{aux}}$ as well as probe detuning $\delta_p$. A set of these 3 parameters define an operating point P, that are chosen after careful optimization. A signal field, $E_{\text{sig}}$ of extremely low-frequency 10 Hz is chosen with amplitude of about 100 mV/cm. Now, the auxiliary modulation amplitude $E_{\text{aux}}$ is scanned from 250 mV/cm to 500 mV/cm while the probe detuning $\delta_p$ is varied from 160 to 270 MHz initially with coarse-grained steps and finally with fine-grained steps. The demodulated output of LIA is recorded at each point and fitted with a sine function. The fitted sine amplitudes are plotted in a 2D color map, which helps to find an optimal operating point. A sample such dataset is shown in Fig.\ref{fig:scanned_grid} with a smaller (zoomed-in) range of $E_{\text{aux}}$ and $\delta_p$ . 
The ``bright yellow'' squares in 2D color map are points where the fitted output signal has maximum amplitude. Therefore, they represent optimal points that can be chosen for sensor operation. In our case, we have chosen $E_{\text{aux}}=354$ mV/cm, $\delta_{p}=243$ MHz, i.e. the square marked with `P'.
\section{Direct low-frequency sensing}
\label{sec:direct_lowfrsens}
\begin{figure}[!h] 
\vspace{-10pt}
	\centering
	\includegraphics[width=0.7\textwidth]{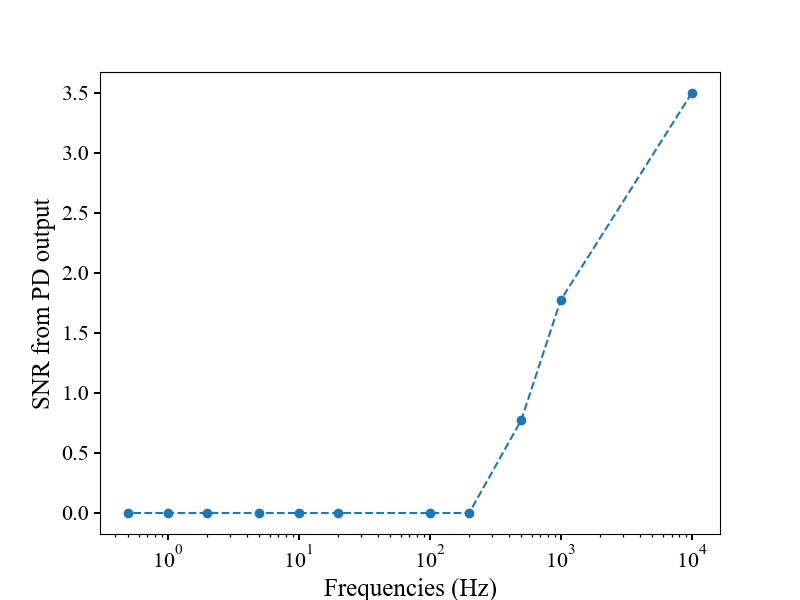} 
	\caption{Direct sensing of low frequencies $\ge500$ Hz with amplitude 100 mV/cm measured at $\delta_p = - 7$ MHz, when E-field applied in vertical direction.}
    \vspace{-6pt} 
	\label{fig:lowf_directsense}
\end{figure}
In this case, the signal field is applied to the vertical pair of copper plates (could be applied to horizontal pair as well), without any auxiliary modulation field.
The frequency to be sensed, appears as modulation on the transmission signal as Rydberg energy level is modulated by the Stark effect.
Fig.~\ref{fig:lowf_directsense} plots the signal-to-noise ratio (SNR) at the output for various frequencies input field of amplitude 100 mV/cm. As shown, frequencies below 500 Hz cannot be detected, and for frequencies $\ge 500$ Hz, SNR improves with increase in signal frequency.
To summarize, direct sensing of low-frequency fields with paraffin-coated cell, does not allow detection of extremely-low frequencies and for low frequencies that are detected, the sensitivity is relatively poor by orders of magnitude when compared to sensitivity obtained from classical receivers.
\section{Demodulated output signals}
\label{sec:demod_out}
\begin{figure*}[!h] 
	\vspace{-8pt}
    \centering
	\includegraphics[width=0.95\textwidth]{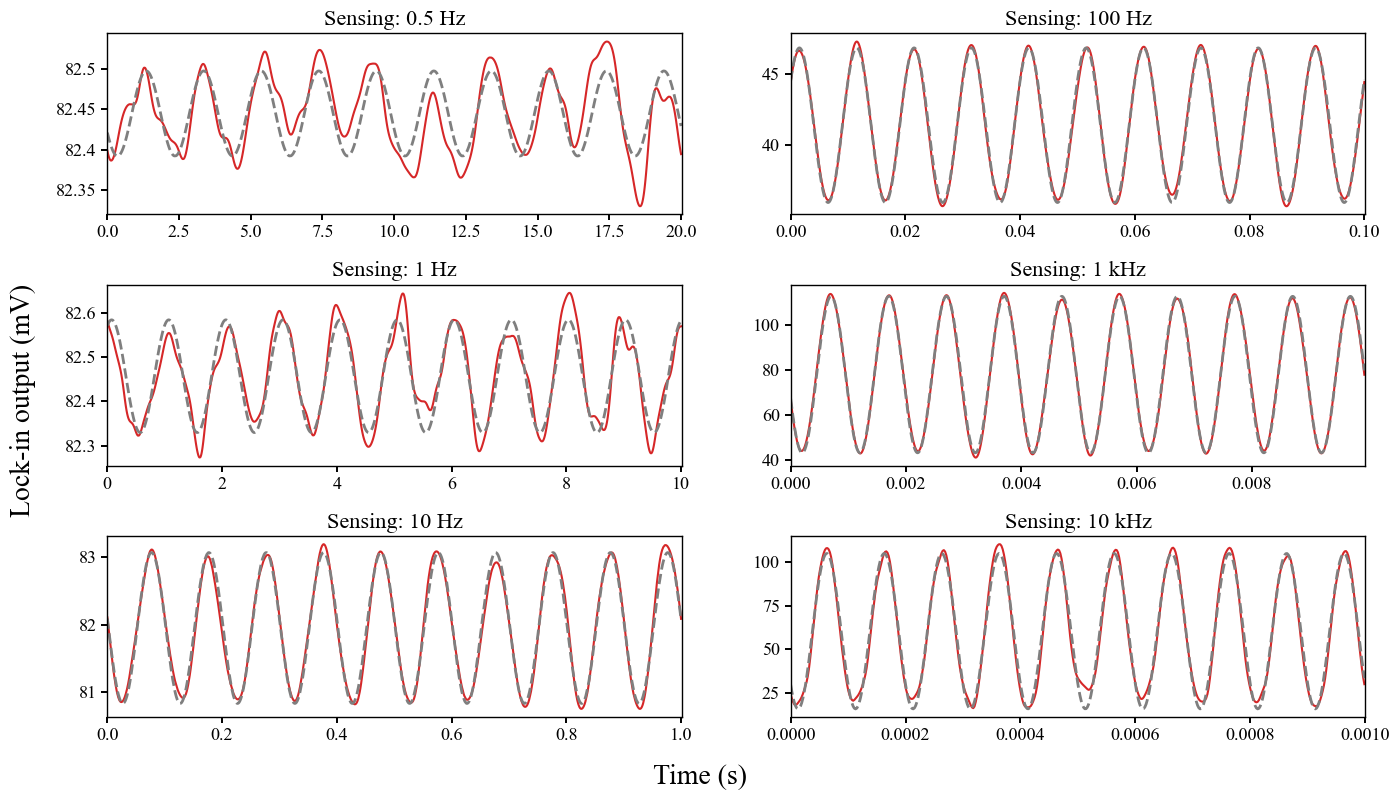} 
	\caption{Demodulated output of lock-in amplifier for frequencies - 0.5 Hz, 1 Hz, 10 Hz, 100 Hz, 1 kHz, 10 kHz with an input (applied) field of 43.7 mV/cm. The dashed lines (gray) are sinusoidal fits to the raw output.}
   \vspace{-8pt}
	\label{fig:lowfreq_sine} 
\end{figure*}
The demodulated raw output of LIA has been plotted in time domain in Fig.~\ref{fig:lowfreq_sine} for frequencies - 0.5 Hz, 1 Hz, 10 Hz, 100 Hz, 1 kHz, 10 kHz with an input (applied) field of 43.7 mV/cm. As seen, the output signals are easily fitted with sine function demonstrating a high signal-to-noise ratio typically achieved in our method. In a similar way, the fitting to the raw output signal has been performed for other values of input amplitudes and 
frequencies.

\section{Noise measurement and analysis}
\label{sec:noise-analysis}
The key to estimation of field sensitivity is to finding out noise sources at a given frequency that fundamentally set the lower limit for detection of E-field amplitude in 1 Hz bandwidth. The impact of various noise sources in our experiment is studied here. The primary sources are - lock-in amplifier (LIA) and balanced photodetector (PD).

\begin{figure}[!t] 
\vspace{-5pt}
	\centering
	\includegraphics[width=0.98\textwidth]{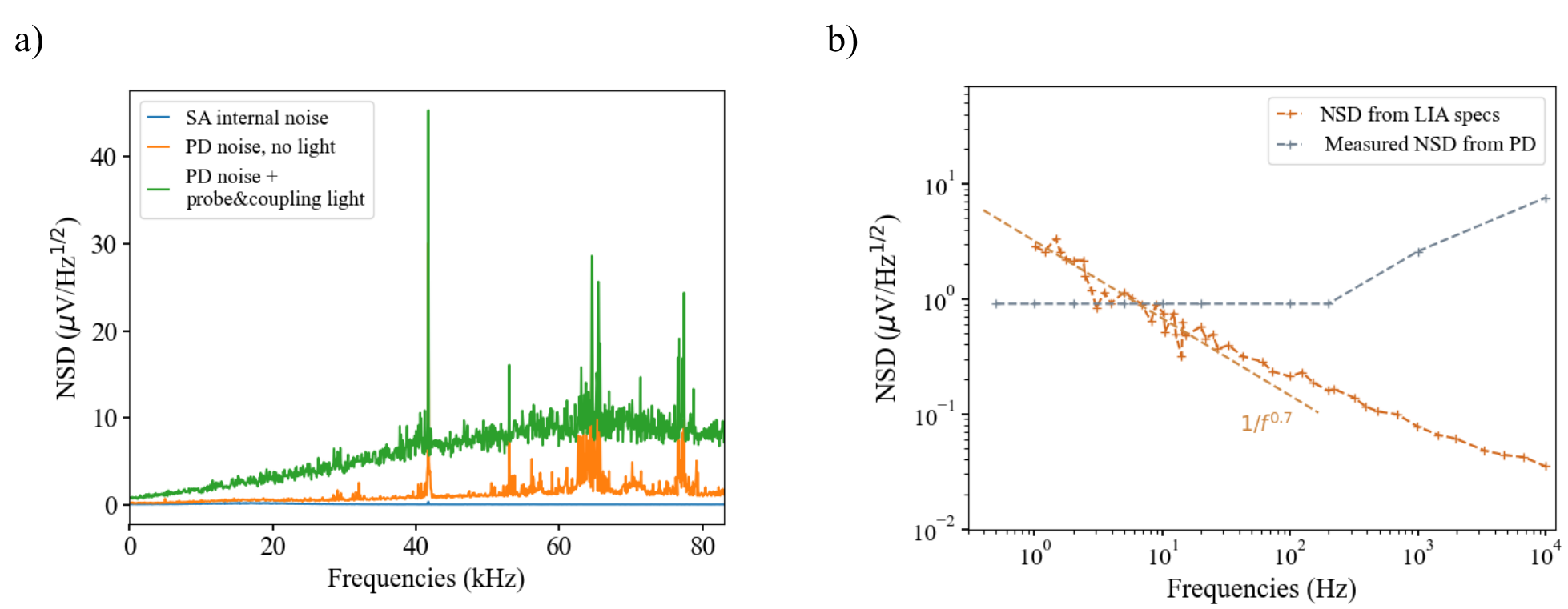} 
	\caption{(a) Noise spectral density (NSD) originating from PD as function of frequency measured via a spectrum analyzer (SA). SA background noise (blue), internal PD noise with no light (orange) and the laser intensity noise combined with PD noise (green) are shown here. (b) Noise spectral density as function of signal frequency showing noise densities from LIA, NSD$_{\text{LIA}}$ (dark gray) and from PD, NSD$_{\text{PD}}$ (brown).}
    \vspace{-5pt}
	\label{fig:PD_noise_meas}
\end{figure}

The balanced photodetector (PD) model is PDB450A from Thorlabs. It has been used with highest gain setting, i.e. trans-impedance gain, G$=10^7$ V/A, bandwidth $=100$ kHz. The total noise from PD is found by measuring the spectral content of PD output via a spectrum analyzer (SA). Fig. \ref{fig:PD_noise_meas}(a) shows such a measurement where the power (dBm converted to $\mu$V) at each frequency is recorded with resolution bandwidth (RBW) of 1 Hz. SA background noise and internal PD noise (with no light)  are shown by blue and orange traces. The green trace captures laser intensity noise combined with PD noise. 
It has been recorded when both lasers are frequency-locked and the sensor in operation (i.e. same probe detuning $\delta_p$ and Rabi frequencies $\Omega_{\text{p}}$, $\Omega_{\text{c}}$ as used in operating point P). Due to lock-in detection, the noise density at modulation frequency $f_{mod}$ is only relevant and the measured values $\text{NSD}_{\text{PD}}= 0.9$, 2.57, 7.54 $\mu$V/$\sqrt{\text{Hz}}$ for $f_{mod}=7.9$, 27.9, 87.9 kHz effectively accounts for all noise originating from PD. They are represented by points (in gray) in Fig.~\ref{fig:PD_noise_meas}(b).
 
The lock-in amplifier (LIA) model is MFLI 500 kHz from Zurich Instruments. Apart from the contribution originating from PD, the inherent noise generated by LIA also adds noise to its demodulated output. This is actually given by the input voltage noise density, that is provided by the device manual. it happens to be flicker noise and when fitted show 1/$f^{0.7}$ dependency. Knowing this dependency helps us to find the noise density at 0.5 Hz which is not provided by the manual.
The relevant values of NSD$_{\text{LIA}}$ are indicated (in brown) in Fig.~\ref{fig:PD_noise_meas}(b).

Now, the total noise spectral density is obtained by combining both spectral densities. Since noise sources are uncorrelated, the total noise spectral density is given by:
$$\text{NSD}_{\text{total}}=\sqrt{\left(\text{NSD}_{\text{PD}}\right)^2 + \left(\text{NSD}_{\text{LIA}}\right)^2}$$ Subsequently, the field sensitivity at signal frequency $f$ is estimated from the relation
$$\text{S}_E = \text{NSD}/m(f)$$
where $m(f)$ is the slope obtained from linear fits to measured data as discussed in the main text. 

For extremely-low frequencies close to DC, the noise floor is typically very high as it is dominated by flicker noise 1$/f^{\alpha}$ where the exponent $0<\alpha<2$ depends on electronic systems (photodetector, ADC, DAC, etc.) under consideration. However in our case, the sensor is unaffected by this noise as lock-in detection technique allows measurements at the modulation frequency $f_{mod}$ in a narrow bandwidth where the noise floor is dominated by the white noise of the PD. Therefore, the narrow detection bandwidth is given by the cutoff frequency (i.e. $1.265f$) of LIA low-pass filter, where $f$ is the signal frequency.
\end{appendices}
\backmatter\\\\
\textbf{Acknowledgements}\\
We would like to thank Alwaleed Aldhafeeri and Dr. Chee Wei Wong for sharing knowledge and valuable insights on laser frequency stabilization. We thank Dr. Fritz Diorico for helpful discussions. We gratefully acknowledge the late Professor Michael Lim of Rowan University, for thoughtful discussions and advice. We remember him with gratitude.\\\\
\textbf{Author contributions}\\
The project was conceived and supervised by RD. AC performed the experiment with NP. AC contributed to the optimization, analysis and wrote the manuscript with inputs from RD. All the authors discussed the results and reviewed the manuscript.\\\\
\textbf{Funding information}\\
The research has been supported by the National Research Foundation (NRF), Singapore and A*STAR under Quantum Engineering Programme (NRF2021-QEP2-03-P01). It has also been supported by NRF, Singapore through the National Quantum Office, under Centre for Quantum Technologies Funding Initiative (S24Q2d0009).\\\\
\textbf{Data availability}\\
All data available from authors upon reasonable request.
\section*{Declarations}
\textbf{Competing interests}\\
The authors have no conflict of interest and no competing interests to declare.


\end{document}